# Predictions for Neutron Scattering and Photoemission Experiments on CuGeO$_3$

Stephan Haas and Elbio Dagotto

*Department of Physics and National High Magnetic Field Lab, Florida State University, Tallahassee, FL 32306, USA*

(August 8, 1995)

Applying numerical techniques to a model recently proposed for the one dimensional (1D) spin-Peierls compound CuGeO$_3$, we calculate dynamical properties that can be directly compared with inelastic neutron scattering (INS) data and angle-resolved photoemission (ARPES) experiments. The momentum and energy dependence of the dynamical structure factor S(q, $\omega$) are discussed, as well as the static structure factor S(q). The spectral function A(q, $\omega$) calculated from the one particle Green's function at half-filling is shown at several values of the hole hopping amplitude t. The results have some unique features characteristic of one dimensional systems including small weight near the Fermi energy. The presence of "shadow bands" induced by short distance antiferromagnetic (AF) correlations is predicted to appear in ARPES experiments for CuGeO$_3$ and also for Sr$_2$CuO$_3$.

64.70.Kb,75.10.Jm,75.50.Ee

The behavior of low dimensional magnetic materials continues to attract much experimental and theoretical interest. A variety of exotic ground states can be observed in different compounds depending on anisotropies, the spin of the magnetic ions, and the coupling to the lattice degrees of freedom. Recently, CuGeO$_3$ was found to be the first inorganic system that shows a spin-Peierls (SP) phase. [1] With decreasing temperature a regime described by uniform S=1/2 Heisenberg chains undergoes a phase transition at T$_{SP}$ = 14K into a dimerized system due to the coupling of the spin-1/2 Cu$^{2+}$-ions to the three-dimensional lattice phonons. At T$_{SP}$ a structural transition takes place where alternate atoms are displaced in opposite directions, and an energy gap for magnetic spin triplet excitations appears. The results of Hase, Terasaki and Uchinokura [1] have generated much excitement, and the presence of a spin-gap in CuGeO$_3$ has been confirmed by several groups. [2,3]

CuGeO$_3$ has a c-direction lattice constant of 2.941Å, much smaller than the a and b lattice parameters. Although the c-axis lattice constant corresponds to the Cu$^{2+}$ − Cu$^{2+}$ distance, the actual links between coppers are provided by the edge sharing of CuO$_6$ octahedra. The angle defined by each Cu-O-Cu bond is $\sim 86°$ (Ref. [4]). This induces an exchange coupling J$_1$ in the effective nearest-neighbors (n.n.) S=1/2 Heisenberg model for the Cu-O chains which is much smaller than those found in the two dimensional (2D) cuprate superconductors. The measured susceptibility, $\chi$, above 14K while showing characteristics of 1D antiferromagnets, is not quantitatively reproduced by calculations corresponding to a n.n. S=1/2 Heisenberg model. Actually, Lorenzo et al. [5] based on a neutron scattering study suggested that a next-nearest-neighbor (n.n.n.) coupling J$_2$ may arise from the Cu-O-O-Cu path. [5] Riera and Dobry [6] arrived to the same conclusion studying numerically the temperature dependence of $\chi$. They reported a substantial ratio J$_2$/J$_1$ = 0.36. In this regime both phonons *and* frustration contribute to the opening of the spin gap. Castilla et al. [7] also remarked the relevance of n.n.n. terms reporting a smaller ratio J$_2$/J$_1$ = 0.24, with J$_1$ = 150K, which is right below the ratio where the effects of frustration open a spin gap in the model in the absence of phonons. Actually, exact diagonalization (ED) techniques locate the critical ratio at J$_2$/J$_1|_c \approx 0.2412 \pm 0.0001$. [8,7] While the microscopic origin of J$_2$ is not quite clear, we include it in our studies simply on phenomenological grounds to model accurately CuGeO$_3$. Since some experimental evidence shows that the SP transition is exclusively driven by phonons, [3] in our study we use Castilla et al.'s estimation of J$_2$. However, note that working so close to J$_2$/J$_1|_c$ likely induces strong dimerizing fluctuations of non-phononic origin in the ground state.

Based on this discussion, the model studied here corresponds to a Heisenberg model with n.n. and n.n.n. interactions and a term that dimerizes the lattice to mimic the effects of phonons. The Hamiltonian is

$$H = J_1 \sum_{\langle ij \rangle} \mathbf{S}_i \cdot \mathbf{S}_j + J_2 \sum_{\langle\langle im \rangle\rangle} \mathbf{S}_i \cdot \mathbf{S}_m + \delta \sum_{\langle ij \rangle} (-1)^i \mathbf{S}_i \cdot \mathbf{S}_j,$$
(1)

where i, j, m denote sites of a 1D chain with N sites, $\mathbf{S}_i$ are S=1/2 spins, and $\langle ij \rangle$ ($\langle\langle im \rangle\rangle$) corresponds to n.n. (n.n.n.) spins along the chain. The static properties of the special case J$_2$ = 0 have been widely studied in the literature, [9] while the frustrated J$_1$ − J$_2$ model has also received much attention. [8] However, *dynamical* studies with controlled approximations have not been previously reported to the best of our knowledge, and here we provide this information using ED techniques. [10]

Inelastic neutron scattering experiments have been recently performed on CuGeO$_3$ single crystals. [14,15] The opening of a gap at q = 0 was observed in the dispersion of the magnetic excitations. An ED study by Castilla et al. [7] showed that the INS triplet dispersion is reproduced using Eq.(1) if $J_2/J_1 = 0.24$ and $\delta = 0.03\,J_1$. We adopt these parameters in the present paper. Thus far the *dynamical* information contained in the energy dependence of $S(q,\omega)$ has not been analyzed in the INS literature for $CuGeO_3$. The reason is that the experimental data is usually contaminated by spurious effects, and a complicated convolution between the actual data and the instrumental resolution is needed to extract $S(q,\omega)$. To guide this experimental effort it is useful to have a *theoretical* prediction for $S(q,\omega)$. The purpose of this paper is to provide for the first time spin dynamical information obtained from model Eq.(1) to compare theory with experiments in $CuGeO_3$. Here we also calculate the spectral function $A(q,\omega)$ which can be directly compared with ARPES experiments. In the context of high temperature superconductors, recent ARPES data for $Sr_2CuO_2Cl_2$ at half-filling has provided important constraints on theoretical models for the cuprates. [11] A similar study in $CuGeO_3$ would likely prove as useful.

chain [13] which has a spin gap caused by a disordered but translationally invariant ground state. In Fig.1b, the first peak in the spectrum is shown as a function of momentum and compared with experiments. [14,15]. As previously remarked by Castilla et al. [7] the agreement between theory and experiment is excellent. From our numerical study we can also analyze the position of the peak with the highest energy to put bounds on the multiparticle continuum. [16] The result is shown in Fig.1b, which in principle could be compared directly with INS results. However, note that the rapidly decaying intensity of the spectral weight makes such comparison difficult. If the resolution of the INS experiment only allows for the observation of intensities within, e.g., 10% or 20% of the maximum at $q = \pi$, then the apparent width of the continuum would appear considerably reduced.

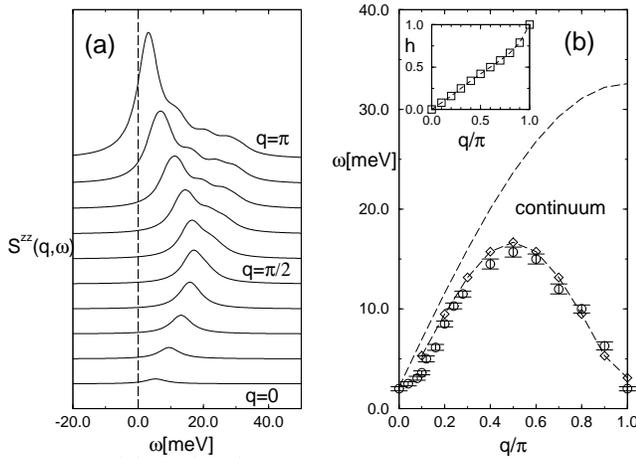

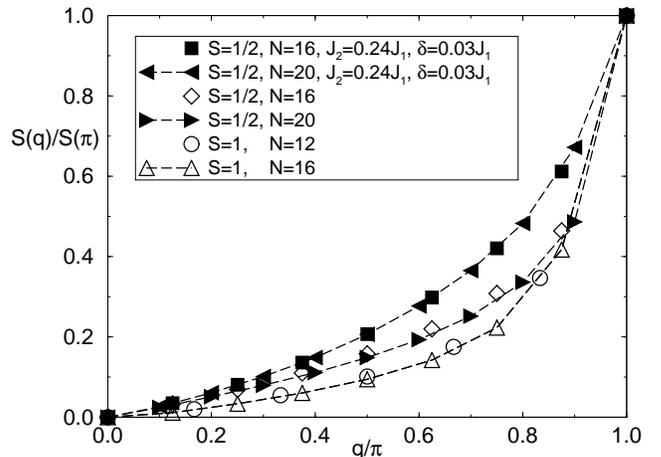

FIG. 1. (a) $S^{zz}(q,\omega)$ obtained on a chain of $N = 20$ sites with a broadening $\epsilon = 0.3 J_1$ at several momenta; (b) Position of the first pole in Fig.1a for each q, defining the spin triplet dispersion (diamonds). The circles denote results from Refs. [14,15]. The dashed line near these points guides the eye, while the dashed line at higher energies signals the position of the pole with the highest energy for each q. In the inset, the intensity h of the highest peak in Fig.1a, normalized with respect to $q = \pi$, is shown as a function of momentum, using $J_2/J_1 = 0.24$, $\delta/J_1 = 0.03$ and $J_1 = 150K$.

In Fig.1a $S^{zz}(q,\omega)$ is shown at several momenta. The position of the dominant peak defines the energy of the first triplet excitation for each q. [12] Note that this peak is not isolated but forms part of a broad continuum. The intensity is the highest at $q = \pi$ as expected due to the power-law decaying AF spin-spin correlations in the ground state. The intensity of the dominant peak for each momentum, normalized with respect to $q = \pi$, is shown in the inset of Fig.1b. This ratio decays away from $q = \pi$ more slowly than in the case of a S=1 Heisenberg

FIG. 2. $S(q)$, normalized with respect to its value at $q = \pi$, as a function of q. The convention for the symbols is shown in the inset. We present results for the model Eq.(1), for the (gapless) n.n. spin-1/2 Heisenberg chain, and for the spin-1 Heisenberg chain which has a disordered ground state and a finite spin gap.

In Fig.2, the static structure factor $S(q)$ ($= \int d\omega\, S(q,\omega)$) is presented. As shown in the figure, finite size effects are small and we consider our results representative of the bulk limit. As expected, the large AF correlations in the model cause $S(q)$ to peak at $q = \pi$, slowly decreasing away from it as the momentum is varied. The normalized results for Eq.(1), with the parameters introduced in Ref. [7], can be clearly distinguished from those obtained at $J_2 = 0$, and thus INS measurements of $S(q)$ can be used to confirm the presence of important n.n.n. couplings in $CuGeO_3$ ($\delta$ is so small that does not influence on $S(q)$ as much as $J_2$). For completeness, results for the S=1 chain are also shown.

We have also calculated the spectral function $A(q,\omega)$ which corresponds to the creation of holes by the removal of spins on the chains in the sudden approximation. As Hamiltonian for this calculation we use the $t-J$ model generalization of Eq.(1) i.e. we add a hopping term

$$H_t = -t \sum_{\langle ij \rangle, \sigma} (\bar{c}^\dagger_{i,\sigma} \bar{c}_{j,\sigma} + \text{h.c.}), \quad (2)$$

where $\bar{c}$ are hole operators and the rest of the notation is standard. [10] In the absence of experiments testing the properties of carriers in $CuGeO_3$, it is difficult to predict the value of the hopping amplitude t. Thus, in the results quoted below three different amplitudes are used such that more information is available to compare with experiments once ARPES data becomes available. However, if the couplings $J_1$ and $J_2$ are caused by an exchange mechanism it is likely that the amplitude t would be larger than both. For simplicity, we have neglected the influence of the dimerization on t which in principle should also be modulated as the exchange $J_1$. Since the dispersive features observed in our calculations have bandwidths of order $J_1$ ($\gg \delta$), the influence of $\delta$ on the ARPES data is expected to be small.

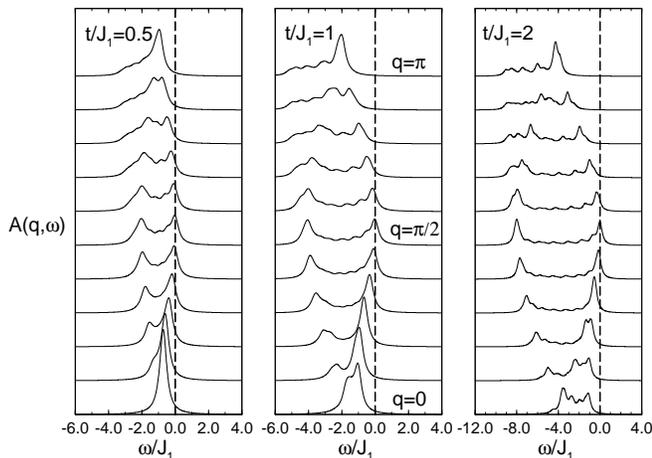

FIG. 3. Spectral function $A(q,\omega)$ on a chain of 20 sites, at several values of $t/J_1$. The broadening is $\epsilon = 0.25 J_1$, and the Fermi energy is arbitrarily located at $\omega = 0$. The other couplings are as in Fig.1.

In Fig.3, $A(q,\omega)$ is shown for different momenta and three couplings $t/J_1$, with $J_1$, $J_2$ and $\delta$ defined as before. Note the presence of a dominant peak at low binding energy which moves towards the Fermi energy (at $\omega = 0$ in the figure) as the momentum is changed from q = 0 to $q = \pi/2$ for all couplings. The lowest binding energy is precisely reached at $q = \pi/2$. Based on previous experience with the 2D copper oxide compounds, [10] this large peak may correspond to a hole immersed in a *locally* ordered AF spin background which increases its effective mass. An interesting feature of our results is the presence of substantial weight at large $\omega$ for all q and couplings. At $q = \pi/2$, the shape of $A(q,\omega)$ at the top and bottom of the band seems similar i.e. a symmetry line seems to run through the middle of the total bandwidth. This feature also appears in the 2D t − J model but only at very small J/t. [17] This is another effect predicted by theoretical calculations that an ARPES experiment could analyze, although we are aware that in such experiments it is difficult to investigate structure at large binding energies due to the presence of core backgrounds in the data that grow with binding energy, as well as the influence of other bands not taken into account in simple one band models like the one used in the present paper.

A curious effect obvious to the eye in Fig.3 is the presence of a robust peak in $A(q,\omega)$ for momenta larger than $q = \pi/2$. This behavior is quite different from what is observed in normal metals, and here we argue that it is caused by the strong AF correlations in the ground state. Recently, this effect, which is usually referred to as "shadow bands", has received much attention in the context of the 2D cuprates after ARPES experiments addressed their existence in Bi2212 at optimal doping. [18] The origin of the shadow bands is simple: in the presence of AF order the unit cell doubles its size due to the nonzero staggered order parameter, reducing the Brillouin zone in half. This induces an extra symmetry in the spectrum since now, e.g., on a square lattice momenta along the diagonal from (0,0) to $(\pi/2,\pi/2)$ become equivalent to those from $(\pi/2,\pi/2)$ to $(\pi,\pi)$. [19] ARPES techniques [11] have shown that antiferromagnetically induced bands can be clearly seen in the insulator $Sr_2CuO_2Cl_2$. If AF order is replaced by a finite but robust AF correlation, the effects of shadow bands are expected to be still visible in ARPES data. The present calculations have shown that shadow bands can be robust *even* in 1D systems, where we are used to the concept that there are no ordered phases at zero temperature. Our prediction can be extended to materials like $Sr_2CuO_3$ where the n.n. Heisenberg model accurately describes its properties. [20] The existence of shadow bands in 1D systems should not be surprising since the hole dispersion tested in ARPES experiments is dominated by *short* distance correlations.

In Fig.4a,b,c, the dispersion of the large features (peaks) observed in Fig.3 are plotted, both at low and high binding energy. The approximate symmetry of the spectrum with respect to the middle of the total bandwidth is clear. In Fig.4d the weight Z of the lowest energy pole appearing in $A(\pi/2,\omega)$ is shown as a function of $N^{-1/2}$ as presented in previous studies for 1D systems. [21] Although Z is very small, it nevertheless seems to converge to a *finite* constant in the bulk even at very large ratios $t/J_1$ contrary to what occurs for the pure t − J model in 1D. [21] Emery [22] recently argued that in the present calculation at *half − filling* with a spin gap in the system, the charge degrees of freedom tested in $A(q,\omega)$ indeed should have a one particle Green's function with a quasiparticle peak carrying a finite Z (see also Ref. [21]). The strong correlations reduce Z to only a fraction of its maximum value 1 (as shown in Fig.4d), but ARPES experiments should still observe a peak with a finite weight in their data. However, as the density of carriers becomes finite, the holes will tend to form a Luttinger liquid with Z=0. The crossover between these two regimes is interesting both theoretically and experimentally if $CuGeO_3$ is doped away from half-filling, and it certainly deserves further study.

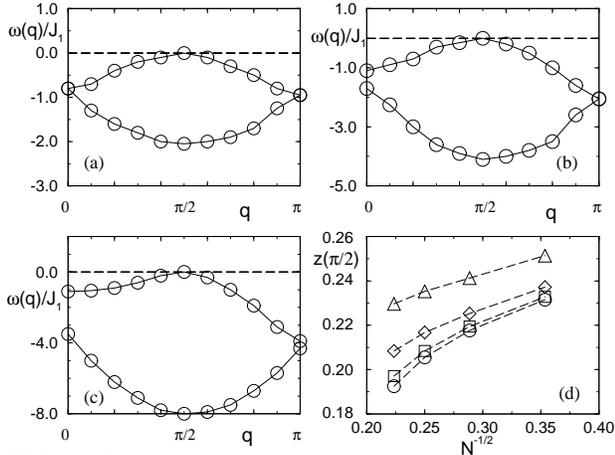

FIG. 4. Dispersion of the poles with the lowest and highest binding energy found in $A(q,\omega)$ using a 20 site chain at different q and for coupling $t/J_1$ equal to (a) 0.5, (b) 1.0 and (c) 2.0. The rest of the couplings are as in Fig.1; (d) Spectral weight $Z(\pi/2)$ of the pole with the lowest binding energy appearing in our calculation at $q = \pi/2$ vs $N^{-1/2}$, where N is the number of sites. The triangles, diamonds, squares and circles correspond to $t/J_1$ equal to 2.0, 8.0, 32.0 and 128.0, respectively. $Z(\pi/2)$ is normalized such that it is bounded between 0 and 1.

Summarizing, here we presented dynamical properties of a model for $CuGeO_3$. Theoretical predictions are made that can be directly compared with experiments. In the context of INS, we have calculated the intensity and momentum dependence of $S(q,\omega)$, as well as the integrated static structure factor $S(q)$. For ARPES experiments at half-filling, the dispersion of quasiparticle-like features was provided. The shape of the spectrum is somewhat unusual with substantial weight concentrated at high energies. We also predict the presence of "shadow bands" in ARPES data induced by AF correlations in the ground state of the 1D compounds $CuGeO_3$ and $Sr_2CuO_3$.

Correspondence with V. J. Emery, L. P. Regnault and Z. X. Shen is acknowledge. We thank D. Duffy and A. Moreo for their comments. E. D. is supported by the Office of Naval Research under grant ONR N00014-93-0495. We also thank the Petroleum Research Fund and MARTECH for additional support.